\documentclass[%
 aip,
 amsmath,amssymb,
 reprint,%
]{revtex4-1}

\usepackage{graphicx}% Include figure files
\usepackage{dcolumn}% Align table columns on decimal point
\usepackage{bm}% bold math
\usepackage[utf8]{inputenc}
\usepackage[T1]{fontenc}
\usepackage{mathptmx}

\begin{document}

\preprint{AIP/123-QED}

\title{Development of an adjustable Kirkpatrick-Baez microscope for laser driven x-ray sources at CLPU}
% Force line breaks with \\

\author{G.Zeraouli}

%\altaffiliation{ 
 %}%\\

\email{ghassan.zeraouli@gmail.com}

\author{G.Gatti}%
\affiliation{ 
	CLPU, Centro de Laseres Pulsados, Building M5, Science Park, Calle Adaja s/n, 37185 Villamayor, Salamanca, SPAIN%\\This line break forced with \textbackslash\textbackslash
}%

\author{A.Longman}

\affiliation{%
	University of Alberta, 116 St  85 Ave, Edmonton, AB T6G 2R3, Alberta, CANADA%\\This line break forced% with \\
}%

\author{J.A.Perez}%
\affiliation{ 
	CLPU, Centro de Laseres Pulsados, Building M5, Science Park, Calle Adaja s/n, 37185 Villamayor, Salamanca, SPAIN%\\This line break forced with \textbackslash\textbackslash
}%
\author{D.Arana}%
\affiliation{ 
	CLPU, Centro de Laseres Pulsados, Building M5, Science Park, Calle Adaja s/n, 37185 Villamayor, Salamanca, SPAIN%\\This line break forced with \textbackslash\textbackslash
}%
\author{D.Batani}%

\affiliation{%
	CELIA, Centre des Laser Intenses et Applications, 351, Cours de la Libération, F-33405 Talence cedex, FRANCE%\\This line break forced% with \\
}%

\author{L.Volpe}%
\affiliation{ 
	CLPU, Centro de Laseres Pulsados, Building M5, Science Park, Calle Adaja s/n, 37185 Villamayor, Salamanca, SPAIN%\\This line break forced with \textbackslash\textbackslash
}%
\affiliation{ 
	University of Salamanca, Salamanca, SPAIN%\\This line break forced with \textbackslash\textbackslash
}%

\author{L.Roso}%
\affiliation{ 
	CLPU, Centro de Laseres Pulsados, Building M5, Science Park, Calle Adaja s/n, 37185 Villamayor, Salamanca, SPAIN%\\This line break forced with \textbackslash\textbackslash
}%
\affiliation{ 
	University of Salamanca, Salamanca, SPAIN%\\This line break forced with \textbackslash\textbackslash
}%

\author{R.Fedosejevs}
%%\homepage{http://www.Second.institution.edu/~Charlie.Author.}
\affiliation{%
	University of Alberta, 116 St  85 Ave, Edmonton, AB T6G 2R3, Alberta, CANADA%\\This line break forced% with \\
}%

\date{\today}% It is always \today, today,

\begin{abstract}
A promising prototype of a highly adjustable Kirkpatrick-Baez (KB) microscope has been designed, built and tested in a number of laser driven x-ray experiments using the high power ($200 TW$) VEGA-2 laser system of the Spanish Centre for Pulsed Lasers (CLPU). The presented KB version consists of two, perpendicularly mounted, $500 \mu m$ thick Silicon wafers, coated with a few tens of $nm$ layer of Platinum unlike the conventional, coated, millimetre thick glass substrates, affording more bending flexibility and large adjustment range. According to simulations, and based on total external reflection, this KB offers a broad-band multi-$keV$ reflection spectra, allowing more spectral tunablity than conventional Bragg crystals. In addition to be vacuum compatible, the prototype is characterised by a relatively small size ($21cm\times31cm\times27cm$) and permits remote control and modification of both the radius of curvature (down to $10 m$) and the grazing incidence angle (up to $60 mrad$). A few examples of focusing performance tests, limitations and experimental campaign results are discussed.
\end{abstract}

\maketitle

\section{\label{sec:level1}Introduction}

Imaging has always been challenging for scientists in all domains. It is one of the first and most important tools used for characterisation in physics, and is usually carried out via visible light transport. On the one hand, imaging resolution is limited by the well-known physical wave property of light, called the diffraction limit. The latter is proportional to the wavelength of the radiation transported, therefore, the smaller the wavelength is, the better the resolution gets. Thus, using x-rays is theoretically the best tool to achieve high resolution imaging. On the other hand, the refractive index of most media used is close to 1 if irradiated with x-rays making the design of refractive imaging components challenging. At the same time the absorption term $\beta$ of the refractive index formula ($n(\omega)=1-\alpha(\omega)-i\beta(\omega)$; where $\alpha$ is the dispersive aspect of the wave-matter interaction) becomes important leading to substantial absorption for lower energy x-rays. To avoid this, one can resort to using ultrathin optics, such as Fresnel zone plates \cite{refzoneplates}.%[$refzoneplates$]
 These are highly efficient in term of transmission and focusing but are highly chromatic\cite{refzoneplateschromatic}%[$refzoneplateschromatic$]
 , which considerably limits the imaging resolution of broadband radiation emitting objects or thermally shifted plasma lines in laser matter interaction experiments. The most widely used approach to overcome this challenge is to use various reflection based optics. One solution is using Bragg imaging crystals, as are frequently used by the Laser-Plasma community for their high reflectivity around the Bragg peack\cite{refBragg}.% [$refBragg$].
  However they have relatively narrow spectral resolution (few tens of $eV$) and manufacturing them is still challenging. In another approach, elliptical graded multilayer x-ray mirrors are considered as one of the best alternatives for optimal x-ray imaging, but, due to their cost, only a few facilities around the world have employed such optics. A practical solution is to use curved, high $Z$ metallic coated mirrors, which are widely used for imaging x-rays. They operate in a grazing incidence configuration and permit an almost complete reflection of the incident x-ray beam below a certain photon energy. We describe in this article, a vacuum compatible Kirkpatrick-Baez (KB) version of such an imaging system designed for multiple laser driven x-ray experiments. The high $Z$ metallic coating of the KB mirrors offer the possibility to operate in a broadband x-ray range in which the spectrally integrated reflectivity can be orders of magnitude higher than Bragg crystals. The computer controlled focusing of the mirrors and alighnment adjustability made it a very flexible diagnostic to carry out x-ray imaging in laser driven plasma experiments.

\section{\label{sec:level1}DIAGNOSTIC DESIGN}
\subsection{\label{sec:level2}Imaging relations}

In the year 1948, the two physicists P. Kirkpatrick and A. Baez designed and invented the first version of the KB microscope \cite{KBinitial}.%[$KBinitial$].
 It was introduced to enable imaging in the x-ray range. Nowadays this device is commonly used for ICF experiments \cite{reffusion}%[$reffusion$]
 (to image the implosion core and fast ignition\cite{reffastignition})%[$reffastignition$])
 and also by the synchrotron community \cite{refsychrotron} .%[$refsychrotron$].  
 We can find the KB in large laser installations like OMEGA\cite{KBOMEGA} and NIF\cite{NIFKB} (USA) and also at LMJ\cite{KBLMJ} (France). %[$KBLMJ$].
  The KB consists of two spherical (or cylindrical) perpendicular mirrors with two different radius of curvature in order to focus the incoming x-rays to a single point. The first mirror focuses the x-rays in the meridional plane and the second mirror focuses the light in the sagittal plane as presented in figure1. 
  
\begin{figure} 
	\includegraphics[scale=0.37]{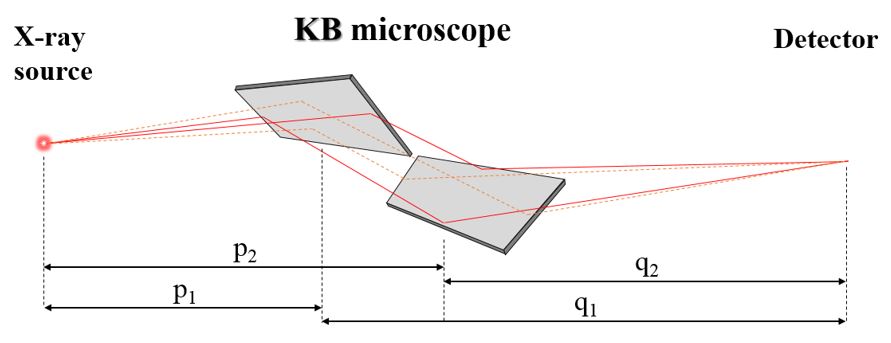}% Here is how to import EPS art
	\caption{\label{fig:epsart} Schematic diagram of a KB microscope system.}
\end{figure}

The physics of the KB is governed by two relations. The first links the focal length to the radius of curvature of the mirror, while the second links the focal length to the imaging distances, object-mirror ($p$) and mirror-image ($q$). We note that the following relations govern the focusing and imaging in the meridional plane for each mirror. 

\begin{eqnarray}
 f_{n} = \frac{ R_{n} \times sin( \theta_{n} )}{2}
\end{eqnarray}
\begin{eqnarray}
\frac{1}{ f_{n} } = \frac{1}{ p_{n} } + \frac{1}{ q_{n} }
\end{eqnarray}

Where $f$  is the focal length of the mirror ($n$ refers to mirror 1 and 2), $\theta$ the grazing incidence angle in ($rad$) and $R$ the radius of curvature of the mirror.

\subsection{\label{sec:level2}Acceptance and working range}

As mentioned before, KB systems are total external reflection based devices, operating at grazing incidence angles and can be optimized for a given x-ray energy range. A high $Z$ material coating is usually used to achieve high reflectivity.
The present KB version was built in such a way that it reflects soft and hard x-rays (up to tens of $keV$) meanwhile it maintains a reasonable ratio (depending on the application) between the number of the incident and reflected photons, and the acceptance angle of the device, i.e, the higher the energy of the incoming photon, the smaller must be the grazing incidence angle, and the lower is the acceptance angle of the KB. Defining $A_{cc} $ as the effective acceptance angle ($rad$) for each mirror of the KB, which is a proportional to the input angle at the entrance of the device and the relative reflectivity at a given grazing angle, we can then write 

\begin{eqnarray}
N_{ref} =F_{inc }\times A_{cc} 
\end{eqnarray}

\begin{eqnarray}
A_{cc} = \int  R_{E} ( \theta ) \times  sin( \theta )\times \frac{dS}{r} 
\end{eqnarray}

Where $F_{inc }$ is the incident x-ray number fluence ($photons/srad$) and Nref is the number of reflected photons per transverse angle, $R_{E} ( \theta )$ is the reflectivity as a function of angle for a chosen energy E at a grazing angle of incidence $\theta$, $S$ is the length coordinate along the illuminated mirror’s surface and $r$ is distance between the x-ray source and the mirror. The total acceptance solid angle $\Omega_{eff}$ is given by 
\begin{eqnarray}
\Omega_{eff} =  A_{cc1}\times A_{cc2}
\end{eqnarray}

Based on the Fresnel model to calculate specular optical reflectivity function, XOP\cite{XOP}(X-ray Oriented Programs) % [$RefXOP$] 
was used to run simulations in order to determine the optimum angle which should be used to reflect the energies wanted. In our approach we use strips of a few hundreds of microns thick ($500 \mu m$) silicon wafers as a mirror substrates instead of conventional millimetre thick glass plates. The use of such thin plates makes it easy to bend and offers a large range for flexible focusing. The strips were coated with a few tens nanometres Platinum layer ($\sim 50 nm$). This choice was due to the high electron density of Platinum which makes it suitable to use at larger angles and higher photon energies (up to a few tens of $keV$). 

\begin{figure}
	\includegraphics[scale=0.35]{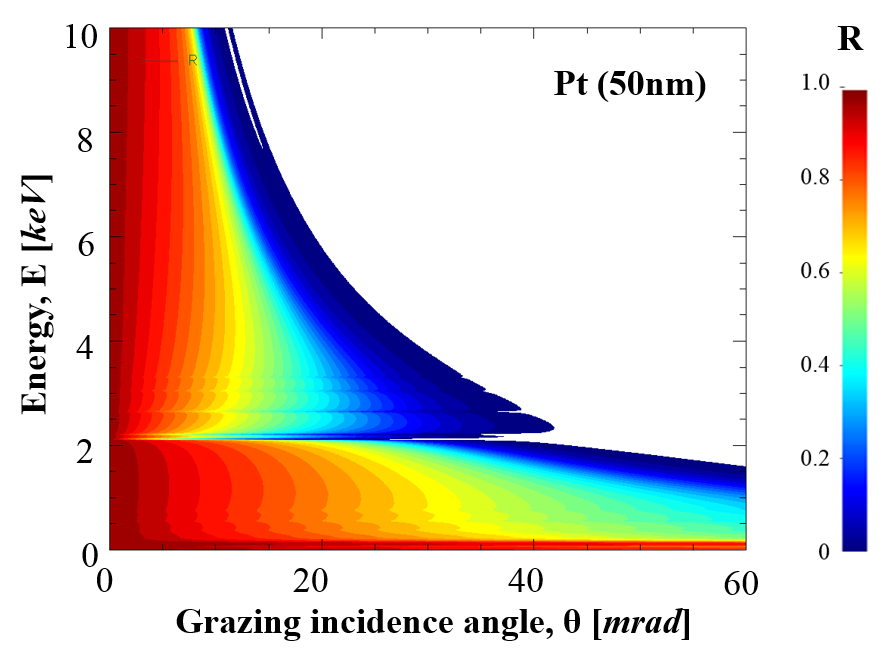}% Here is how to import EPS art
	\caption{\label{fig:epsart} Simulated reflectivity map of a $50 nm$ thin Pt layer in function of energy and grazing angle of incidence.}
\end{figure}

\begin{figure}
	\includegraphics[scale=0.59]{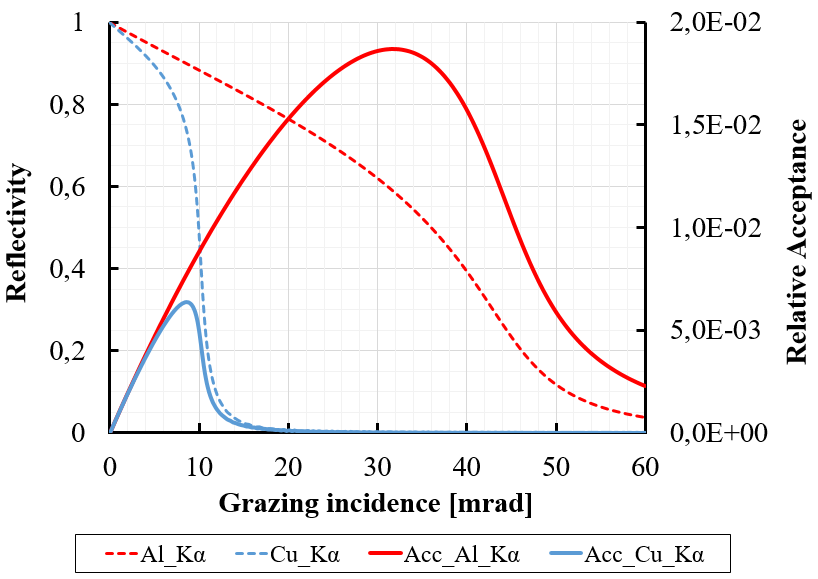}% Here is how to import EPS art
	\caption{\label{fig:epsart} Reflectivity functions of $Al K\alpha $ and $Cu K\alpha $ on a single Pt coated mirror and corresponding product function by the acceptance angle at each grazing incident angle.}
\end{figure}

The KB mechanical assembly consists of two independent benders remotely controlled by two piezoelectric motors each. In order to achieve bending, two different coupled forces should be applied to the edges of the mirrors. For simplicity, and based on simulations, we have chosen to use the ‘cantilever’ bending mechanism, discribed in ref \cite{Bendingtheory}.%[$Bendingtheory$]. 
 ($Newfocus8301$) Piezo-electric actuators were used to generate the couple of forces required to bend the mirrors. These motors were chosen for the relative high axial force they can apply ($20 N$) and for the small step size ($30 nm$) which offers more remote controlling accuracy. This system has also manually adjustable lateral twisting arms which can be used to remove any residual twist in the mirror plates. Laboratory tests have demonstrated the importance of these two arms in both guiding and shaping the beam (circles in Fig.4).

\begin{figure}
	\includegraphics[scale=0.4]{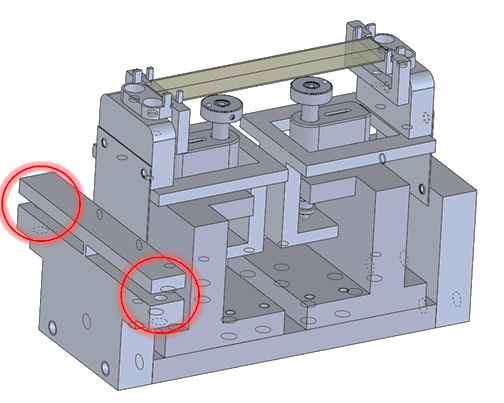}% Here is how to import EPS art
	\caption{\label{fig:epsart} Scheme of an isolated bender in which the twisting arms can be seen (in red).}
\end{figure}

\begin{figure}
	\includegraphics[scale=0.23]{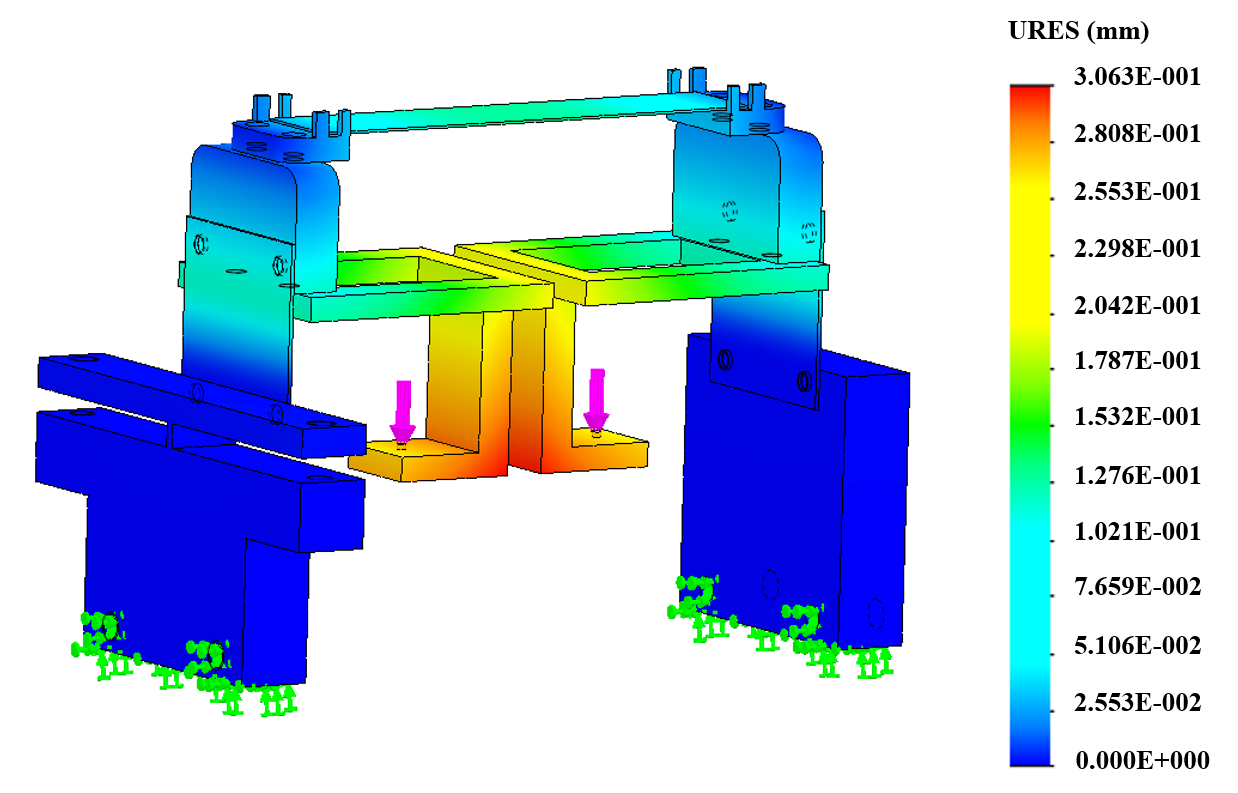}% Here is how to import EPS art
	\caption{\label{fig:epsart} Displacement simulation of the elements of a signle bender after applying a $20N$ force (pink vectors) on the pushing arms.}
\end{figure}

\begin{figure}
	\includegraphics[scale=1.4]{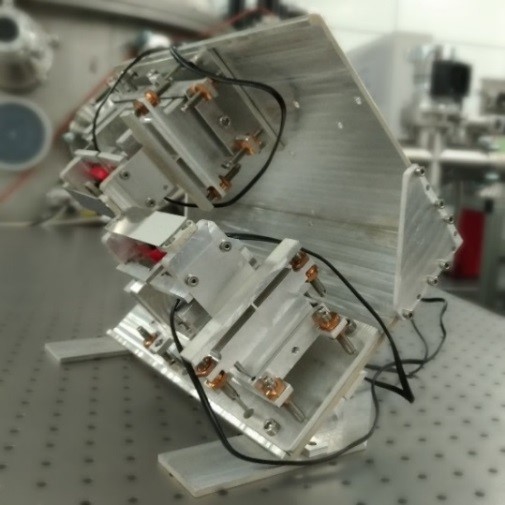}% Here is how to import EPS art
	\caption{\label{fig:epsart} Mounted Kirkpatrick-Baez microscope in CLPU’s target area.}
\end{figure}

The single bender mechanism design scheme, a stress simulation example and the final 3D design of our x-ray imager are shown in figures $4$, $5$ and $6$ respectively. The benders were oriented at $45$ degree angles, relative to horizontal, as shown in Fig.6 in order that the total angular deflection of the x-ray beam remained in the horizontal plane so that the final detector could be mounted in the horizontal plane of the target system. By assembling the mirror assemblies, a critical step is the attachments of the mirror strips to the bender mechanism. In this case, the mirrors were glued in place using epoxy glue with the bender in its relaxed state.

\section{\label{sec:level1}OPTICAL MEASUREMENT TESTS}
\subsection{\label{sec:level2}Slope error theory and measurements}

The first study we have carried out was meant to determine the surface shape of our silicon strip mirrors and then allow us to simulate the predicted resultant x-ray image spots using ray tracing software. The mirror strips were cut out of standard high grade $500\mu m$ thick Silicon $4$ inches diameter wafers. We have used total reflection based optical measurements to achieve optimum results. The experimental setup consists of expanding a $0.5 mm$ He-Ne laser beam to a $10 cm$ diameter beam truncated by a $2 cm$ wide slit as a probe beam. Then each bender was placed, independently, in a near normal incidence reflection mode (see Fig.7). For each mirror, a camera was placed at a given position of $13.3 m$ from the bender, a bit farther than the exact focal distance. 

\begin{figure}
	\includegraphics[scale=0.31]{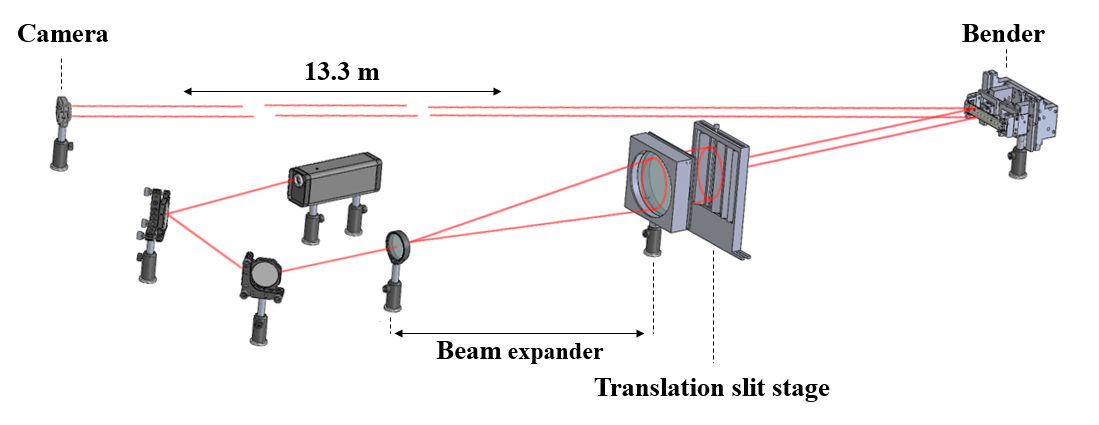}% Here is how to import EPS art
	\caption{\label{fig:epsart} Parameter definition for angular deviation calculation.}
\end{figure}

\begin{figure}
	\includegraphics[scale=0.26]{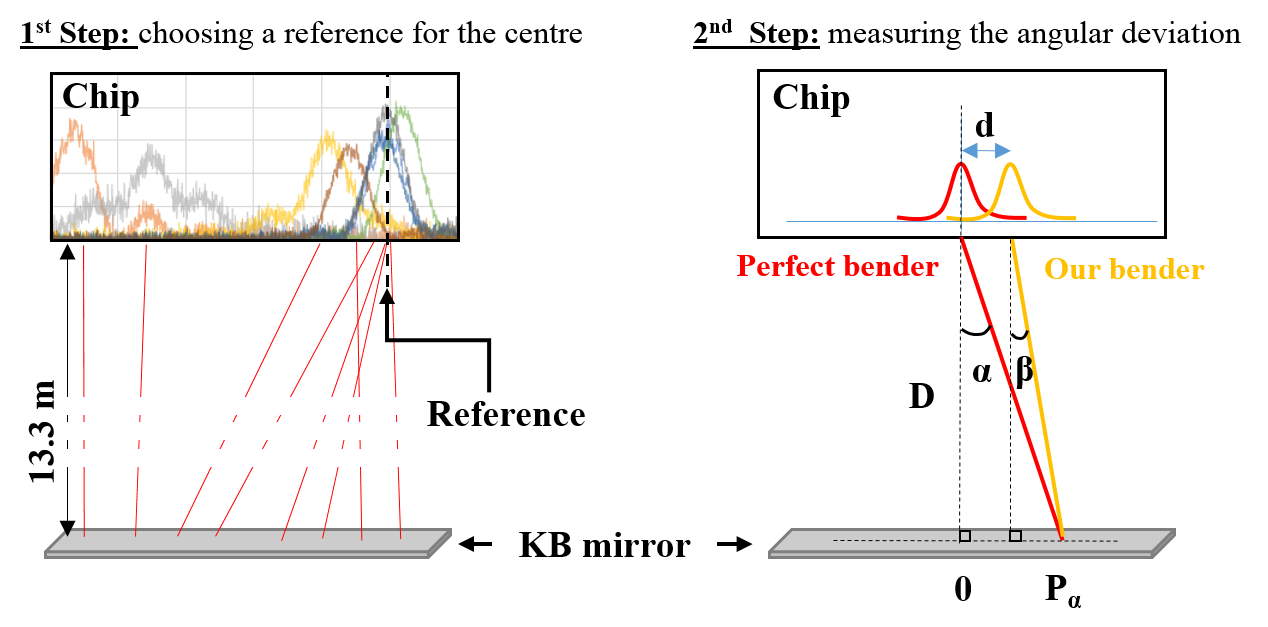}% Here is how to import EPS art
	\caption{\label{fig:epsart} Experimental setup for surface shape characterisation.}
\end{figure}

Using to a translation stage, we were able to move the slit situated before the mirror, allowing the independent measurement of each region of the mirror. The incident signal was collected by a camera and then analysed following the steps in figure 8. Slope error determination was carried out by measuring the angular deviation technique, which consists on reconstructing the mirror’s shape depending on the angle of incidence from each region of the mirror (see Fig.9).

\begin{figure*}
\includegraphics[scale=0.5]{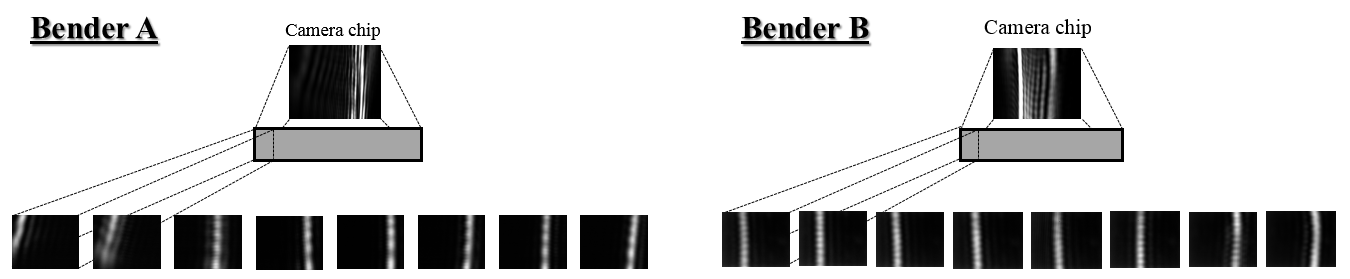}% Here is how to import EPS art
\caption{\label{fig:wide}(Bottom) Experimental results of the focus of isolated mirror segments on the camera for both mirror benders. (Top) Sum of all rays on the camera (when the translation slit is fully open).}
\end{figure*}

From the data, one can clearly see that each mirror has slightly different deflection patterns corresponding to its specific shape which is due to variations within the standard manufacturing tolerances of the silicon wafers. 
According to Fig.8, the angular deviation can be written as following.

\begin{eqnarray}
\mid   \Psi \mid = \big( \frac{ \alpha-  \beta }{2} \big) 
\end{eqnarray}

Where $\alpha$ and $\beta$ are the angles shown in the Fig.8, measured from the focus position of a theoretical perfect bender and our measured bender, respectively.
\begin{eqnarray}
 \alpha = tan^{-1}  \big( \frac{ P_{ \alpha } }{D} \big) 
\end{eqnarray}
\begin{eqnarray}
\beta = tan^{-1}  \big( \frac{ P_{ \alpha }-d}{D} \big) 
\end{eqnarray}

Where $P_{i}$ refers to the position of the incident beam on the mirror’s surface, $D$ the distance from the camera to the mirror and $d$ the separation between the real and the theoretical image position on the plane of the camera. Using the measured results, the unknown focal length is found by minimizing the rms average value of $d$ for the set of measurements. Then the remaining angular deviation can be integrated as a function of position to give residual error in surface height. After calculation, the final results are shown in the Fig.10, in which the letters A and B refers to the two mirrors in the KB system, and the numbers 1 and 2 refer to the edges of each mirror when the incident beam hits edge 1 first. It can be seen that the mirrors made from standard high grade Silicon wafers have a surface accuracy of around $1\mu m$.

\begin{figure}
	\includegraphics[scale=0.45]{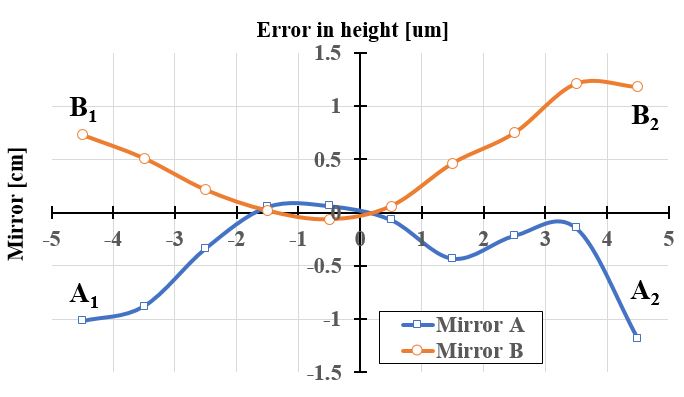}% Here is how to import EPS art
	\caption{\label{fig:epsart} Reconstructed shape result of mirror A and mirror B.}
\end{figure}

\subsection{\label{sec:level2}Focal spot analysis}
A second study was carried out to characterize the imaging accuracy of the KB microscope. We installed the two benders into the KB microscope system, placed it into the expanded beam path of a collimated alignment laser and then, remotely adjusted its focus into a given camera to obtain the smallest spot size. The experimental setup is shown in Fig.11. The grazing incidence for this measurement was set at 2 degrees.

\begin{figure}
	\includegraphics[scale=0.35]{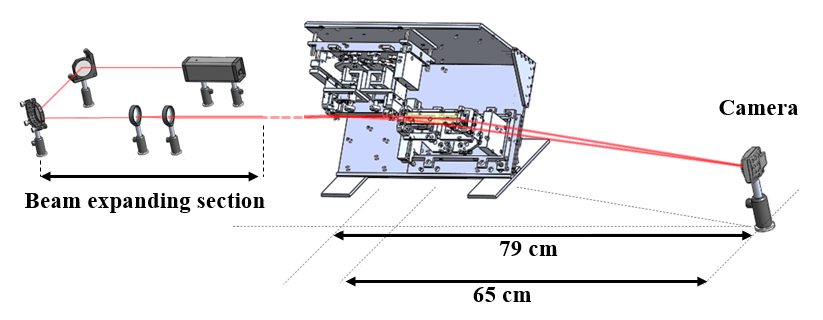}% Here is how to import EPS art
	\caption{\label{fig:epsart} Experimental setup for focal spot optimisation.}
\end{figure}

The aperture of the KB is rectangular, so, one can calculate the reflection spot using the Fraunhofer diffraction formula for a slit aperture, in each transverse direction, given by:

\begin{eqnarray}
I(\theta)= I_{0}sinc^{2}( \frac{d \pi }{ \lambda } sin( \theta ))
\end{eqnarray}

Where $d$ is the mirror’s effective aperture size, $\lambda$ wavelength of the incoming beam, $\theta$ the angle of deviation off axis. Using the small angle approximation, we replace the term sin($\theta$) by $x/f$, where $f$ will be the focal length and $x$ the lateral displacement from the centre axis on the plane of the detector.

\begin{eqnarray}
I(x)= sinc^{2}( \frac{d  \pi x }{ \lambda f } )
\end{eqnarray}

The measured focal spot is typically larger than expected from the diffraction limit, corresponding to the whole mirror surface width. From the measurement, one can extract the parameter $d$. This effective size gives an estimate of how much of the mirror length is contributing to the reflected spot. This allows for a calculation of the mirror efficiency, $\eta$, which then can be calculated by dividing the aperture size $d$ of the experimental fitted signal to the theoretical one. 

\begin{eqnarray}
 \eta_{i}= \frac{ d_{i,exp} }{d _{i,Th} } 
\end{eqnarray}

Where the index $i$ can be A or B and refers to the benders. The total estimated efficiency is then given by

\begin{eqnarray}
 \eta_{KB}= \eta_{A} \times  \eta_{B}= \frac{ d_{A,exp} }{d _{A,Th} } \times\frac{ d_{B,exp} }{d _{B,Th} }
 \end{eqnarray}

 Figure 12 shows the focal spot obtained together with the plotted profiles for both benders. The experimental focal spot size (FWHM) obtained using a He-Ne laser beam ($633 nm$) for the given distances is $130.8\mu m$ and $108.5 \mu m$ for the benders A and B respectively, meanwhile the theoretical expected values are $126.94 \mu m$ and $104.44 \mu m$. According to the results of both, experiment and simulation shown in Fig.12, $\eta_{KB}$ was equal to 93\%.
 
 \begin{figure}
 	\includegraphics[scale=0.45]{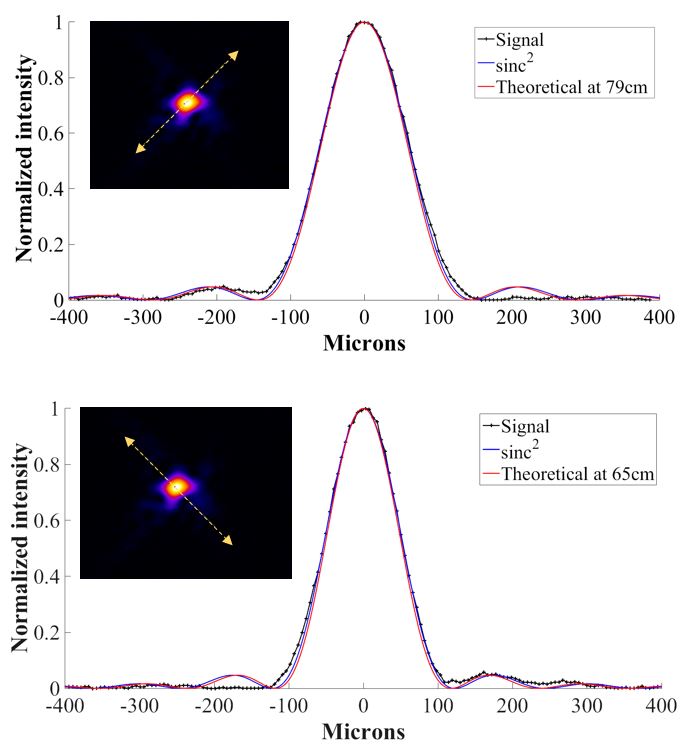}% Here is how to import EPS art
 	\caption{\label{fig:epsart} Focal spot profile of mirror A (top) and mirror B (bottom).}
 \end{figure}
 
 \section{\label{sec:level1}Application}
 
 The KB microscope has been used in an initial experimental campaign on the CLPU VEGA2 ($200 TW$) Laser system to study proton acceleration via the TNSA mechanism. The KB microscope was used to image $Al K\alpha$ radiation generated from laser-driven hot electron transport into a $6 \mu m$ thick Al foil target. The collection of these radiation has been carried out by an x-ray CCD camera (Greateyes model $GE 1024 256$) with $26\mu m$ pixel size. The latter was covered with two layers of $2\mu m$ of Mylar coated with thin layers of a $100 nm$ Aluminium for blocking the visible light. An additional $6\mu m$ Aluminium foil was added in the KB’s path to block low energy x-ray photons. The remotely controllable KB was placed $1.2 m$ away from the target center at $30$ degrees with respect to the target normal and adjusted in such a way to create an image of the x-ray radiation source exactly in the detector’s plane. To enable direct comparison to optical measurements, the distances from the KB to the camera were maintained to be the same as for optical measurements as given in Fig.$11$. We used a shot to shot method to focus and optimise gradually the KB by remote control. The results of the focusing procedure and the final focus image obtained are shown in figures 13 and 14. This setup was also used for carrying out x-ray radiography measurements, in which we placed a metallic grid of $110 \mu m$ crossing wires with $110 \mu m$ open spaces in the x-ray beam path with the results shown in Fig.14.  
 
\begin{figure}
	\includegraphics[scale=0.25]{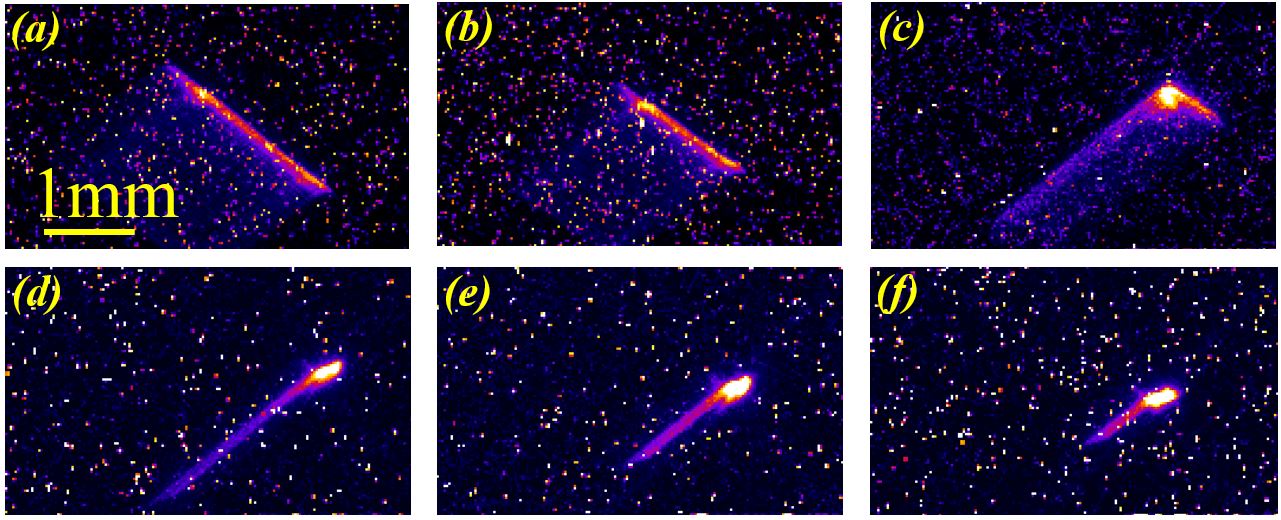}% Here is how to import EPS art
	\caption{\label{fig:epsart} Carried out focusing steps to achieve final KB imaging.}
\end{figure}

\begin{figure}
	\includegraphics[scale=0.75]{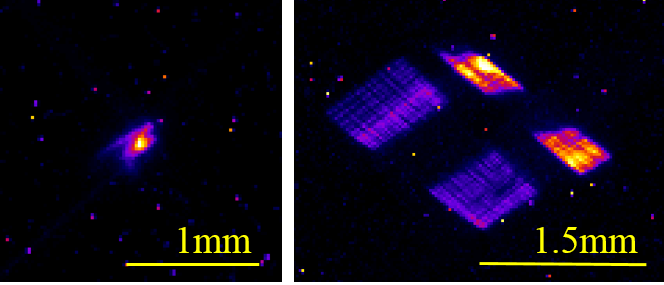}% Here is how to import EPS art
	\caption{\label{fig:epsart} Final result of an adjusted KB’s focus (left). Magnified x-ray radiography projection of a metallic grid (right).}
\end{figure}

In in Fig.13 we can see how the focal spot of the KB is adjusted from shot to shot with real $K\alpha $ radiation signal. The first three pictures show the remote controlling effect, of the exclusive bending of one mirror, on shaping the focus, meanwhile the other three show the effect of the other bender. The final measured spot size of the KB was $134.2\mu m$ by $99.6\mu m$ for the mirrors A and B respectively (see Fig.14.left). A magnified x-ray radiography projection has been achieved too during the experiment as shown in Fig.14.right. In this case the demagnification ratio for the imaging system was $M_{A}\sim 0.66$ and $M_{B}\sim 0.54$, where the indexes A and B refer to the benders. The final image size observed here is a convolution of the actual $Al K\alpha$ spot size due to lateral spread of the refluxing hot electrons and the image spot size due to the imperfection of the mirror surface. In addition, the resolution of this KB image was limited by the x-ray CCD pixel size of $26\mu m$. As seen from Fig.10, the mirror surfaces are approximately an order of magnitude less flat than polished mirror plates usually employed in conventional KB microscopes. This surface height variation would lead to an image spot size of the order of $50$ to $100 \mu m$  and thus could be a significant contributing factor to the final measured spot size. The fact the measured optical spot shown in fig.12 was very close to the predicted values would indicate that contribution of surface aberrations for the final optimized KB microscope are significantly less than a spot size of $100 \mu m$. 

\section{\label{sec:level1}Discussion and conclusions}

The main advantage of our KB design is its flexibility and low cost. As such it offers an important alternative for an area of application in x-ray imaging and x-ray beam transport for laser-plasma experiments where flexibility in setup is important for varying experimental layouts. The final resolution of the current system only depends on the quality of the mirrors allowing the user to choose more expensive polished mirrors if they require resolution down to microns. The mechanical simulation calculations as shown in Fig.5 indicate that perfectly flat silicon plates will bend to curved surface that is accurate within $0.1 \mu m$, allowing for high resolution systems. Thus we believe the simple and flexible KB microscope design presented here should have wide application areas in x-ray transport and imaging for laser-plasma and laser-fusion related studies.

\bigskip

\begin{acknowledgments}

Authors acknowledge to FURIAM project FIS2013-4774-R, PALMA project FIS2016-81056-R, Junta de Castilla y León project CLP087U16 and LaserLab Europe (EU-H2020 654148). A.Longman and R.Fedosejevs would like to acknowledge funding from an NSRC Canada research grant number RGPIN 2014-05736.	
	
\end{acknowledgments}

\end{document}